\documentclass[11pt,a4paper]{article}
\usepackage{amsmath,amsfonts,amssymb}
\usepackage[latin1]{inputenc}
\usepackage[numbers]{natbib}
\usepackage{authblk}
\usepackage[dvips]{graphicx}
\usepackage{array}
\usepackage{comment}
\usepackage{authblk}
\usepackage{color}
\usepackage{multirow}
\usepackage{psfrag}
\usepackage{soul}
\usepackage{enumerate}
\usepackage{caption}
\usepackage[normalem]{ulem}
\usepackage{authblk}
\usepackage{authblk}
\usepackage{amsmath,amssymb}
\usepackage[dvips]{graphicx}
\usepackage{array}
\usepackage{comment}
\usepackage{authblk}
\usepackage{color}
\usepackage{psfrag}
\usepackage{soul}
\usepackage{rotating}
\usepackage{multirow}
\usepackage{enumerate}
\usepackage{caption}
\usepackage[normalem]{ulem}

\newcommand{\fig}[1]{Figure~\ref{#1}}
\newcommand{\equ}[1]{(\ref{#1})}

\newcommand{\ud}{\mbox{d}}

\title{The Effect of Autocorrelation on the Shewhart-RZ Control Chart}
\author[1]{H. D. Nguyen}
\author[2]{A. Ahmadi Nadi}
\author[2]{K.P. Tran\thanks{kim-phuc.tran@ensait.fr (corresponding
    author)}}
\author[3]{P. Castagliola}
\author[4]{G. Celano}
\author[1]{K.D. Tran}

\affil[1]{Institute of Artificial Intelligence and Data Science, Dong A University, Danang,
Vietnam.}
  \affil[2]{Universit\'e de Lille, ENSAIT, GEMTEX,
	F-59000 Lille, France.}

  \affil[3]{Universit\'e de Nantes \& LS2N UMR CNRS 6004, Nantes, France}
  \affil[4]{University of Catania, Department of Civil Engineering and Architecture,
  Catania, Italy.}

\begin{document}
\setlength{\parindent}{0mm}
\pagestyle{plain}
\maketitle

\begin{abstract}
In many industrial manufacturing processes, the quality of products depends on the relation between two main ingredients or characteristics. Often, this calls for monitoring the ratio of two normal random variables with statistical process control (SPC) techniques. A large number of studies related to designing control charts monitoring this ratio has been published. However, these studies are based purely on the assumption of independent observations. 
In practice, autocorrelation between observations can exist and should be modelled to protect against the false alarm rate inflation when implementing a control charts. In this paper, we tackle this problem by investigating the performance of the Shewhart control chart monitoring the ratio of two normal variables, (denoted as Shewhart-RZ), in the presence of autocorrelation between successive observations. The autocorrelation is modelled through the bivariate autoregressive model VAR(1). We also provide an example to illustrates the use of
the Shewhart-RZ control chart on a quality control problem. 
\end{abstract}


\section{Introduction}
In industrial manufacturing practice, control charts are a powerful tool to reduce variability and achieve process stability. Many control charts have been designed in SPC literature to monitor the stability of many parameters characterizing the distribution of a quality characteristic like the Mean, the Median, and the Coefficient of Variation (CV). Recently, the ratio between two normal random variables $X$ and $Y$ has been considered as the quality characteristic to be monitored. The reason is that the ratio can play an important role in ensuring the product quality of various processes. A number of situations where monitoring the ratio of two  variables is needed has been broadly discussed in \citet{Celano2014b} and \citet{Tran_2018_compositional}. Other examples related to the need to monitoring the ratio can be found in \citet{Spisak1990} and \citet{Davis1991}, from an unemployment insurance quality control
program, or  \citet{Oksoy1994} from on-line
monitoring implemented in the glass industry. In particular, the statistical properties of a Shewhart-RZ control chart have been presented in  \citet{Celano2014a} (for individual measurements, i.e by assuming the sample size $n=1$) and \citet{Celano2014b} (for multiple measuremenst, i.e when the sample size $n>1$)). Then, other advanced control charts have been proposed, such as the Synthetic-RZ control chart (\citet{Celano2015}), the Run Rules-RZ control chart (\cite{Tran2015a}), the EWMA-RZ (Exponentially Weighted Moving Average) control chart (\citet{Tran2015b}),  the CUSUM-RZ (Cumulative Sum) control chart  (\citet{Tran2016a}), the VSI EWMA-RZ (Variable Sampling Interval) control chart (\cite{Nguyen_VSIEWMARZ_2018}). Also, the effect of measurement error on the ratio control chart has  been investigated in \citet{Tran2016b} by using
a linear covariate error model. The authors showed that the performance of
this control chart is significantly affected by the presence of
measurement errors. \\

All the control charts monitoring the ratio have been designed under the assumption that the observations collected for $X$ and $Y$ are independent. This assumption overlooks the problem of autocorrelation among consecutive observations, which is very frequent when measures are collected at high rate. Modelling the consecutive observations as time series is very important when correlation exists to avoid negative effects on the performance of the control chart monitoring the process. The first study in SPC literature that deals with autocorrelation  was  conducted by \citet{Alwan1988} who illustrated statistical modeling and fitting of time-series 
effects and the application of standard control charts. From 
that pioneering work, many
studies concerning the performance of various control charts in the
presence of autocorrelation as well as the ways
to handle this autocorrelation have been proposed. \citet{Kalgonda2004} showed that the autocorrelation has a serious impact on the performance of conventional control charts.
The autoregressive moving average control chart  was considered in \citet{Lin2012}. The first-order autoregressive model (AR(1)) was applied to describe the
wandering behavior of the process mean using the $\bar{X}$ chart in  \citet{Franco2012}.  \citet{Leoni2015a} studied the effect of the autocorrelation on the performance of the $T^2$ chart. A synthetic chart to control bivariate processes with autocorrelated data was designed in \citet{Leoni2015b}. Other studies related to  control charts monitoring the processes with autocorrelated observations are in 
\citet{Hwarng2010}, \citet{costa2011}, \citet{Huang2013},
\citet{Huang2014}, \citet{Franco2014a}, \citet{Franco2014b}, for example. \\

 As far as we know, up to now, the effect of the autocorrelation on the
Shewhart-RZ control chart has not been yet considered in the literature. Therefore, the
goal of this paper is to investigate the performance of the 
Shewhart control chart for monitoring the ratio of two normal
random variables that, similar to \citet{Leoni2015a}, are modelled by means of the first-order bivariate VAR(1) model to describe the autocorrelation between observations from the process. \\

The
paper is organised as follows:  in Section \ref{sec:autoregressive}, 
the multivariate autoregressive model is discussed for the application to the sample ratio; in Section 
\ref{sec:RZchart}, the formulas for the control limits and the performance 
metrics of the Shewhart-RZ control chart are discussed; in Section 
\ref{sec:effect}, the effect of the autocorrelation on the Shewhart-RZ 
control chart performance is investigated; in Section \ref{sec:illustrative}, an illustrative
example is provided to show the implementation of the Shewhart-RZ
control chart in the presence of autocorrelation; some concluding remarks and recommendations for future research are given in
Section \ref{sec:conclusions}. The ratio distribution is also provided in the Appendix.
\section{Autoregressive model for the sample ratio}
\label{sec:autoregressive}

Let us assume that, at time $i=1,2,\ldots$, the samples 
$\{\mathbf{W}_{i,j}=(X_{i,j},Y_{i,j})\}$, $j=1,2,\ldots, n$  of size $n$ are collected to monitor the ratio between the normal random variables $X$ and $Y$. Autocorrelation between $X$ and $Y$ is modelled by the bivariate VAR(1) autoregressive model. According to this time series model, $\mathbf{W}_{i,j}$
depends on $\mathbf{W}_{i,j-1}$ through the following equation
\begin{equation}
\label{equ:VAR1}
\mathbf{W}_{i,j}=\boldsymbol{\mu}_{\mathbf W}+\boldsymbol{\Phi}(\mathbf{W}_{i,j-1}
-\boldsymbol{\mu}_{\mathbf W})+\boldsymbol{\varepsilon}_{i,j},
\end{equation}

where
\[
\boldsymbol{\mu}_{\mathbf W}=\left(\begin{array}{c} \mu_X \\
                                     \mu_Y \end{array}\right)
\]

is the mean vector of $\mathbf{W}_{i,j}$, 
\[
\boldsymbol{\Phi}=\left(
  \begin{array}{cc}
    \Phi_{11} & \Phi_{12} \\
    \Phi_{21} &  \Phi_{22}
  \end{array}\right)
\]

is a $(2\times 2)$ correlation matrix that accounts for both autocorrelation, ($\Phi_{i,i}$, for $i=1,2$), and cross-correlation, ($\Phi_{i,j}$, for $i \neq j$), between $X$ and $Y$; 
$\boldsymbol{\varepsilon}_{i,j}$ is an independent bivariate normal random
vector with mean vector
$\boldsymbol{\mu}_{\boldsymbol{\varepsilon}}=\mathbf{0}$ and
variance-covariance matrix
\[
\boldsymbol{\Sigma}_{\boldsymbol{\varepsilon}}=\left(
  \begin{array}{cc}
    \sigma_{eX}^2 & \sigma_{eXY} \\
    \sigma_{eXY} & \sigma_{eY}^2
  \end{array}\right).
\]

From \equ{equ:VAR1}, \citet{Reinsel2003} showed that   the variance-covariance matrix
$\boldsymbol{\Sigma}_{\mathbf W}$ of $\mathbf{W}_{i,j}$ is the solution of the equation 
$\boldsymbol{\Sigma}_{\mathbf W}=\boldsymbol{\Phi}
\boldsymbol{\Sigma}_{\mathbf W}\boldsymbol{\Phi}^{\intercal}
+\boldsymbol{\Sigma}_{\boldsymbol{\varepsilon}}$.
Then, it can be shown that 
\begin{equation}
\label{equ:VecSigmaW}
\mathrm{Vec}(\boldsymbol{\Sigma}_{\mathbf W})=
(\mathbf{I}_{4}-\boldsymbol{\Phi}\otimes\boldsymbol{\Phi})^{-1}
\mathrm{Vec}(\boldsymbol{\Sigma}_{\boldsymbol{\varepsilon}}),
\end{equation}

{where $\mathbf{I}_{p}$ is the $ p\times p $ identity matrix}, $\otimes$ is
the Kronecker product and $\mathrm{Vec}$ is the operator that
transforms a matrix into a one-column vector by stacking its columns.   { Using    \eqref{equ:VecSigmaW}, the variance-covariance matrix of   the considered VAR(1)  autoregressive model in \eqref{equ:VAR1} can be obtained as
\begin{align}\label{equ:SigmaW}
\boldsymbol{\Sigma}_{\mathbf W}&=\begin{pmatrix}
\sigma^{2}_{X} & \sigma_{XY}\\
\sigma_{XY} & \sigma^{2}_{Y}
\end{pmatrix}\nonumber\\
&=\begin{pmatrix}
\frac{\Delta_{11} \sigma_{eX}^2+ (\Delta_{12}+\Delta_{13}) \sigma_{eXY}+ \Delta_{14}   \sigma_{eY}^2}{\Delta} & \frac{\Delta_{31} \sigma_{eX}^2+ (\Delta_{32}+\Delta_{33}) \sigma_{eXY}+ \Delta_{34}   \sigma_{eY}^2}{\Delta} \\
\frac{\Delta_{21} \sigma_{eX}^2+ (\Delta_{22}+\Delta_{23}) \sigma_{eXY}+ \Delta_{24}   \sigma_{eY}^2 }{\Delta}
& \frac{\Delta_{41} \sigma_{eX}^2+ (\Delta_{42}+\Delta_{43}) \sigma_{eXY}+ \Delta_{44}   \sigma_{eY}^2}{\Delta} \\
\end{pmatrix}
\end{align}
where the quantities $\Delta_{ij}$ for $i,j=1,2,3,4$ are given in Appendix 2. \\

Now, one needs to calculate  mean vector $\boldsymbol{\mu}_{\overline{\mathbf W}}$  and  variance-covariance matrix $\boldsymbol{\Sigma}_{\overline{\mathbf W}}$  of the  sample mean vector $\mathbf{\overline W}_i=(\bar X_i,\bar Y_i)^{\intercal}$   at time $i=1,2,\ldots$. It is proven in Appendix 3 that   $\boldsymbol{\mu}_{\overline{\mathbf W}}=\boldsymbol{\mu}_{\mathbf W}=(\mu_X,\mu_Y)^{\intercal}$ and
\begin{align}\label{equ:VCwbar}
\boldsymbol{\Sigma}_{\overline{\textbf{W}}}&=\frac{1}{n} \Bigg[ \boldsymbol{\Sigma}_{\textbf{W}} \left(\mathbf{I}_{2}+\Lambda(\boldsymbol{\Phi}^{\intercal})-\frac{1}{n}\Pi(\boldsymbol{\Phi}^{\intercal})\right)+\left(\Lambda(\boldsymbol{\Phi})-\frac{1}{n}\Pi(\boldsymbol{\Phi}) \right) \boldsymbol{\Sigma}_{\textbf{W}}^{\intercal}\Bigg] \nonumber\\
&=\begin{bmatrix}
\sigma^{2}_{\bar{X}} & \sigma_{\bar{X},\bar{Y}}\\
\sigma_{\bar{X},\bar{Y}} & \sigma^{2}_{\bar{Y}}
\end{bmatrix} 
\end{align}
where 
\begin{align*} 
\Lambda(\boldsymbol{\Phi})&=(\boldsymbol{\Phi}-\boldsymbol{\Phi}^{n})(\mathbf{I}_{2}-\boldsymbol{\Phi})^{-1}\\ 
\Pi(\boldsymbol{\Phi})&=(\boldsymbol{\Phi}^{-1}-\mathbf{I}_{2})^{-1}\Big((\mathbf{I}_{2}-\boldsymbol{\Phi}^{n-1})(\mathbf{I}_{2}-\boldsymbol{\Phi})^{-1}-(n-1)\boldsymbol{\Phi}^{n-1}\Big).
\end{align*}
 
 From \eqref{equ:VCwbar}, it is straightforward to obtain the coefficients of
variations $\gamma_{\bar X}=\frac{\sigma_{\bar X}}{\mu_X}$ and
$\gamma_{\bar Y}=\frac{\sigma_{\bar Y}}{\mu_Y}$ of $\bar X_i$ and $\bar Y_i$. Similarly, the coefficient of correlation
 between $\bar X_i$ and $\bar Y_i$ and   the standard-deviation ratio can be calculated as $ \bar{\rho}=\frac{\sigma_{\bar{X},\bar{Y}}}{\sigma_{\bar{X}}\sigma_{\bar{Y}}}$ and 
$\bar\omega=\frac{\sigma_{\bar X}}{\sigma_{\bar Y}}$, respectively.  Concerning $\bar\omega$ it is important to note that, if we define $z=\frac{\mu_X}{\mu_Y}$, it can also be rewritten as
\begin{equation}\label{equ:omegabar}
\bar\omega=\frac{\gamma_{\bar X}}{\gamma_{\bar Y}}\times\frac{\mu_X}{\mu_Y}
=\frac{\gamma_{\bar X}}{\gamma_{\bar Y}}\times z.
\end{equation}
}\\
{
In the rest of the paper, we will assume that  $X$ and $Y$ are not cross-correlated and the matrix
$\boldsymbol{\Phi}$ is diagonal,
i.e. $\Phi_{12}=\Phi_{21}=0$. This assumption allows reducing the
bivariate stationary conditions to the more simple univariate stationary
conditions for each variable and it has already been stated by \citet{Leoni2015a}, who did they have not found any example
in the SPC literature where the observations of one variable at a time $i, i=2,3,\ldots$ are dependent of the observations of the other variable at
time $i-1$. As proved in \citet{Leoni2015a}, if
$\Phi_{12}=\Phi_{21}=0$ then, using \equ{equ:VecSigmaW}, the
variance-covariance matrix $\boldsymbol{\Sigma}_{\mathbf W}$ of
$\mathbf{W}_{i,j}$ is equal to

\begin{eqnarray}
\label{equ:cova}
\boldsymbol{\Sigma}_{\mathbf W}&=& \left(
  \begin{array}{cc}
    \sigma_{X}^2 & \rho\sigma_{X}\sigma_{Y} \\
    \rho\sigma_{X}\sigma_{Y}  &  \sigma_Y^2
  \end{array}
\right)\nonumber \\
&=&
\left(
  \begin{array}{cc}
   (1-\Phi_{11}^2)^{-1}  \sigma_{eX}^2 &(1-\Phi_{11}\Phi_{22})^{-1} \sigma_{eXY} \\
    (1-\Phi_{11}\Phi_{22})^{-1} \sigma_{eXY} & (1-\Phi_{22}^2)^{-1} \sigma_{eY}^2
  \end{array}
\right),
\end{eqnarray}
where $\sigma_X$ and $\sigma_Y$ are the standard deviations of two components $X$ and $Y$ of $\mathbf{W}$, respectively, and $\rho \in (-1, 1)$ is the corresponding coefficient of correlation.

At time $i=1,2,\ldots$, the sample mean vector
$\mathbf{\overline W}_i=(\bar X_i,\bar Y_i)$  is also a bivariate
normal random vector with mean vector $\boldsymbol{\mu}_{\mathbf W}$
and variance-covariance matrix
\[
\boldsymbol{\Sigma}_{\mathbf{\overline W}}=
\left(
  \begin{array}{cc}
    \xi_{11} & \xi_{12}  \\
    \xi_{21} & \xi_{22}
  \end{array}
\right),
\]

where
\begin{eqnarray*}
  \xi_{11} & = & \frac{\sigma_X^2}{n}\left(1+\frac{2}{n}\sum_{k=1}^{n-1}(n-k)\Phi_{11}^k\right), \\
  \xi_{22} & = &\frac{\sigma_Y^2}{{n}}\left(1+\frac{2}{n}\sum_{k=1}^{n-1}(n-k)\Phi_{22}^k\right), \\
  \xi_{12} & = & \frac{\sigma_{XY}}{n}\left(1+\frac{1}{n}\sum_{k=1}^{n-1}(n-k)\Phi_{11}^k+\frac{1}{n}\sum_{k=1}^{n-1}(n-k)\Phi_{22}^k\right),\\
  \xi_{21} & = &\xi_{12}.
\end{eqnarray*}

From these quantities, it is straightforward to obtain the coefficients of
variations $\gamma_{\bar X}=\frac{\sqrt{\xi_{11}}}{\mu_X}$ and
$\gamma_{\bar Y}=\frac{\sqrt{\xi_{22}}}{\mu_Y}$ of $\bar X_i$ and
$\bar Y_i$ as
\begin{eqnarray}
\label{equ:gammabarX}
\gamma_{\bar X} & = & \frac{\sigma_X\sqrt{1+\frac{2}{n}\sum_{k=1}^{n-1}(n-k)\Phi_{11}^k}}{\sqrt{n}\mu_X}, \\
\label{equ:gammabarY}
  \gamma_{\bar Y} & = & \frac{\sigma_Y\sqrt{1+\frac{2}{n}\sum_{k=1}^{n-1}(n-k)\Phi_{22}^k}}{\sqrt{n}\mu_Y},
\end{eqnarray}

Similarly, the coefficient of correlation
$\bar\rho=\frac{\xi_{12}}{\sqrt{\xi_{11}\xi_{22}}}$ between $\bar X_i$
and $\bar Y_i$ is defined as
\begin{equation}
\label{equ:rhobar}
  \bar\rho=\frac{\rho\left(1+\frac{1}{n}\sum_{k=1}^{n-1}(n-k)\Phi_{11}^k+\frac{1}{n}\sum_{k=1}^{n-1}(n-k)\Phi_{22}^k\right)}
  {\sqrt{\left(1+\frac{2}{n}\sum_{k=1}^{n-1}(n-k)\Phi_{11}^k\right)\left(1+\frac{2}{n}\sum_{k=1}^{n-1}(n-k)\Phi_{22}^k\right)}},
\end{equation}

and the standard-deviation ratio
$\bar\omega=\sqrt{\frac{\xi_{11}}{\xi_{22}}}$ is
\begin{equation}
\bar\omega=\frac{\sigma_X}{\sigma_Y}\times\sqrt{\frac{1+\frac{2}{n}\sum_{k=1}^{n-1}(n-k)\Phi_{11}^k}{{1+\frac{2}{n}\sum_{k=1}^{n-1}(n-k)\Phi_{22}^k}}}.
\end{equation}

Concerning $\bar\omega$ it is important to note that, if we define
$z=\frac{\mu_X}{\mu_Y}$, it can also be rewritten as
\begin{equation}
\label{equ:omegabar}
\bar\omega=\frac{\gamma_{\bar X}}{\gamma_{\bar Y}}\times\frac{\mu_X}{\mu_Y}
=\frac{\gamma_{\bar X}}{\gamma_{\bar Y}}\times z.
\end{equation}

}

{
\section{Estimating the VAR(1) model}

In practice, the  VAR(1) model parameters should be estimated from the phase I data. Several methods have been presented in the literature to obtain these estimated parameters such as the     least squares (LS),    maximum likelihood (ML), and Bayesian    methods.   \citet{Tsay2013}  (Chapt. 2)  provided a detailed discussion on the theoretical aspects of these  estimation methods and studied their key  properties.  They showed that the Bayesian   and   LS methods produce close estimates and  under some  conditions,    the ML estimates  asymptotically equivalent to the LS ones. They also   prepared R demonstrations for each method and  included all the programs  in the \texttt{MTS} package available in R.  \texttt{MTS}   is a general package for analyzing multivariate linear time series. Other statistical software such  SAS can also be used to perform estimation of  VAR models. \\

Here, as an example,  we use   the command  \texttt{VAR}  in \texttt{MTS} package    that uses   LS  method  to   estimate the model of a real dataset.   Let consider the bivariate time series data of size 186 that describes the measured pressure at the front and back end of a furnace. The data set is reported in    \citet{Montgomery2008} (p.  347).  They showed that VAR(1) model provides an appropriate fit for the data.    Based on the data set, we have
 $\boldsymbol{\mu}_{\mathbf{W}}=\boldsymbol{\mu}_{\bar{\mathbf{W}}}=(\bar{W}_{F},\bar{W}_{B})^{\intercal}=(10.421,20.189)^{\intercal}$ where   $W_{F}$ and $W_{B}$ are the pressures at the front and back end of the furnace, respectively. The  output of \texttt{VAR} command shows the estimated VAR(1) autoregressive model is
 \begin{align*}
\begin{pmatrix}
W_{i,F}\\
W_{i,B}
\end{pmatrix}=
\begin{pmatrix}
10.421\\
20.189
\end{pmatrix}+\begin{pmatrix}
0.733 & 0.474\\
0.410 &  -0.561
\end{pmatrix}\left[ \begin{pmatrix}
W_{i-1,F}\\
W_{i-1,B}
\end{pmatrix}-
\begin{pmatrix}
10.421\\
20.189
\end{pmatrix} \right]+\begin{pmatrix}
\varepsilon_{i,F}\\
\varepsilon_{i,B}
\end{pmatrix}
\end{align*} 
where 
\begin{align*}
\begin{pmatrix}
\varepsilon_{i,F}\\
\varepsilon_{i,B}
\end{pmatrix} \sim N_{2}\left(\textbf{0},\begin{pmatrix}
1.232&0.588\\
0.588 & 1.072 
\end{pmatrix}\right).
\end{align*}

It can be seen that these estimated parameters are very close to those of SAS ARIMA procedure    in \citet{Montgomery2008}.   From \eqref{equ:SigmaW} and \eqref{equ:VCwbar}, we obtained
\begin{align*}
\boldsymbol{\Sigma}_{\mathbf W}&=\begin{pmatrix}
5.887  & 1.500 \\
1.500 &2.002  
\end{pmatrix}\\
\boldsymbol{\Sigma}_{\overline{\mathbf{W}}}&=\begin{pmatrix}
4.724 & 1.458 \\
1.458& 0.542 \\
\end{pmatrix},
\end{align*}
where $\boldsymbol{\Sigma}_{\overline{\mathbf{W}}}$ is calculated for $n=5$. Finally, one can see that
\begin{align*}
 \gamma_{\bar{X}}&=\frac{\sigma_{\bar{X}}}{\mu_{X}}=\frac{\sqrt{4.724}}{10.421}=0.209, \quad
 \gamma_{\bar{Y}}=\frac{\sigma_{\bar{Y}}}{\mu_{Y}}=\frac{\sqrt{0.542}}{20.189}=0.036,\\
 & \bar{\rho}=\frac{\sigma_{\bar{X},\bar{Y}}}{\sigma_{\bar{X}}\sigma_{\bar{Y}}}=\frac{1.458}{\sqrt{4.724}\sqrt{0.542}}=0.911, \quad \bar{\omega}=\frac{\gamma_{\bar{X}}}{\gamma_{\bar{Y}}} \times z_{0}=\frac{ 0.209}{0.036} \times \frac{10.421}{20.189}= 2.996.
\end{align*}

}

\section{Implementation of a Shewhart-RZ control chart for autocorrelated data}
\label{sec:RZchart}
Similar to \citet{Tran2015a}, we suggest to monitor the ratio statistic  as
\begin{equation}
  \bar Z_i=\frac{\bar X_i}{\bar Y_i}=\frac{\sum_{j=1}^n X_{i,j}}
  {\sum_{j=1}^n Y_{i,j}}
\end{equation}
at each time $i=1,2,\ldots$.

Let $\rho=\rho_0$ and  $z=\frac{\mu_X}{\mu_Y}=z_0$ denote the coefficient of correlation between the two normal variables $X_{i,j}$ and $Y_{i,j}$ and the mean ratio when the process is in-control, respectively. As in the design of any control charts, the Shewhart-RZ chart for autocorrelated data is designed by defining its control limits. 
It is well-known that the distribution of the ratio of two normal random variables has no moment. Thus, the control limits  of the Shewhart-RZ control chart monitoring $\bar Z_i$ should be defined as probability control limits. Denoting as $\alpha$ the desired false alarm probability for the control
chart, the  lower control limit ($LCL$) and the upper control limit ($UCL$) of the Shewhart-RZ are obtained as follows:
\begin{eqnarray}
  \label{equ:LCLRZ}
  LCL & = & F_Z^{-1}\left(\left.\textstyle\frac{\alpha}{2}\right|
            \gamma_{\bar{X}},\gamma_{\bar{Y}},\bar\omega,\bar\rho\right), \\
  \label{equ:UCLRZ}
  UCL & = & F_Z^{-1}\left(\left.1-\textstyle\frac{\alpha}{2}\right|
            \gamma_{\bar{X}},\gamma_{\bar{Y}},\bar\omega,\bar\rho\right),
\end{eqnarray}

where  $\gamma_{\bar{X}}$, $\gamma_{\bar{Y}}$, $\bar\omega$,
$\bar\rho$ are computed using equations
\equ{equ:gammabarX}--\equ{equ:omegabar} with $\rho=\rho_0$ and
$z=z_0$, and $F_Z^{-1}(\dots)$ is the inverse cumulative distribution
function for the ratio of two normal random variables. An
approximation for $F_Z^{-1}(\dots)$ is given in the Appendix {\color{red} 1}. \\
 

When the process runs in the out-of-control state, the in-control ratio $z_0$ is shifted to $z_1=\tau\times z_0$, where $\tau>0$ is the shift size, and the in-control coefficient of correlation $\rho=\rho_0$ is shifted to $\rho=\rho_1$. According to the discussion in
\citet{Leoni2015a}, we can assume that the units within a sample are
collected close together in time and, at the same time, the length $h$
of the sampling interval is large enough to eliminate any dependence
between successive values of $\bar Z_1,\bar Z_2,\ldots$. That is to say, the observations in each subgroup of size $n$ are autocorrelated but not the statistics $Z_i, i=1,2,\ldots$. As a result, the run length of the
Shewhart-RZ control chart for autocorrelated data  follows a geometric distribution of parameter $1-\beta$, where 
\begin{equation}
\label{equ:beta}
\beta=F_Z\left(UCL\left|\gamma_{\bar{X}},\gamma_{\bar{Y}},\bar\omega,\bar\rho
  \right.\right)-F_Z\left(LCL\left|\gamma_{\bar{X}},\gamma_{\bar{Y}},\bar\omega,
    \bar\rho\right.\right),
\end{equation}
where $\beta$ denotes the
probability of the event that the control chart does not trigger any alarm to show the occurrence of an
assignable cause given that this cause does exist. The out-of-control parameters $\gamma_{\bar{X}}$, $\gamma_{\bar{Y}}$, $\bar\omega$, $\bar\rho$
in (\ref{equ:beta}) are calculated based on equations
\equ{equ:gammabarX}--\equ{equ:omegabar} but with $\rho=\rho_1$ and
$z=z_1$. Then, the out-of-control average run length of the Shewhart-RZ chart is computed as
\begin{equation}
  \label{equ:ARL}
  ARL_1=\frac{1}{1-\beta}.
\end{equation}
\section{The effect of autocorrelation on the Shewhart-RZ Control chart}
\label{sec:effect}
In this Section, we present the effect of autocorrelation on the performance of the Shewhart-RZ control evaluated by using  the average run length ($ARL$). When the
process is in-control, the $ARL$ will be denoted as
$ARL_0$.
The $ARL$ values are computed from (\ref{equ:ARL}) given the fixed values of parameters $n$, $\gamma_X=\frac{\sigma_X}{\mu_X}$, $\gamma_Y=\frac{\sigma_Y}{\mu_Y}$, 
$\Phi_{11}$, $\Phi_{22}$, $z_0$, $\rho_0$, $\rho_1$ and $\tau$.
 Without loss of generality, we assume that $z_0=1$.  { We also suppose    that  $X$ and $Y$ are not cross-correlated and the matrix $\boldsymbol{\Phi}$ is diagonal, i.e. $\Phi_{12}=\Phi_{21}=0$. It allows reducing the bivariate stationary conditions to the more simple univariate stationary conditions for each variable and it has already been stated by \citet{Leoni2015a}. This assumption means that  $X_{i,j} (Y_{i,j})$ and $Y_{i,j-1} (X_{i,j-1})$   are not linearly dependent.}  The in-control $ARL$ is set at $ARL_0=200$, corresponding to the false alarming rate $\alpha=0.005$. In this study, we also assume $n\in\{2,5,7,10,15\}$, $\gamma_X\in\{0.01,0.2\}$, $\gamma_Y\in\{0.01,0.2\}$,  $\rho_0\in\{-0.8,-0.4,0,0.4,0.8\}$, and $\Phi_{11},\Phi_{22}\in\{0.1,0.7\}$. \\

The values control limits $(LCL,UCL)$ of the Shewhart-RZ control chart for the case $\Phi_{11}$= $\Phi_{22}=0.01$ are presented in Table \ref{tab:LUCL}. Results have also been obtained for other parameters but they are not presented here and are available upon request from authors. Similar to the Shewhart-RZ control chart for independent observations investigated in \citet{Celano2014b}, it can be seen from this table that with the same value of sample size $n$, the values of $LCL$ and $LCL$ depend on both the value of $\rho_0$, (given $(\gamma_X,\gamma_Y)$), and the value of $(\gamma_X,\gamma_Y)$, (given the value of $\rho_0$). The increase of $\rho_0$ leads to an increase of $LCL$ and a decrease of $UCL$. For example, with 
  $(\gamma_X,\gamma_Y)=(0.01,0.01)$ 
  and $n=5$, we have $LCL=0.9725$ and $UCL=1.0283$ when $\rho=-0.8$ , compared to  $LCL=0.9840$ and $UCL=1.0163$ when $\rho_0=0.4$. By contrast, the increase of $(\gamma_X,\gamma_Y)$ leads to the decrease of $LCL$ and the increase of $UCL$. For example,  with $n=7$ and $\rho_0=0.0$, we 
  have $LCL=0.9823$ and $UCL=1.0180$ when 
  $(\gamma_X,\gamma_Y)=(0.01,0.01)$ 
  compared to  $LCL=0.6933$ and $UCL=1.4423$ when 
  $(\gamma_X,\gamma_Y)=(0.2,0.2)$.

 \begin{center}
INSERT TABLE \ref{tab:LUCL} ABOUT HERE
\end{center}

We present the effect of the autocorrelation between observations on the Shewhart-RZ chart in Tables \ref{tab:ARL1122}-\ref{tab:ARL1122rphi} assuming that the correlation coefficient $\rho$ between the random variables $X$ and $Y$  dose not be affected by the assignable causes, i.e $\rho_0=\rho_1$. In particular, Tables \ref{tab:ARL1122}-\ref{tab:ARL1122r} show the values of the $ARL_1$ for the case $\Phi_{11} = \Phi_{22}$ and { Tables  \ref{tab:ARL1122Phi}-\ref{tab:ARL1122rphi}} show the values of the $ARL_1$ when $\Phi_{11} \neq \Phi_{22}$. In general, the obtained results show that given the values of other parameters, the increase of $\Phi_{11}$ or $\Phi_{22}$, or both of them leads to the increase of the $ARL_1$; the larger the values of $(\Phi_{11}, \Phi_{22})$, the larger the values of the $ARL_1$. For example, in Table~\ref{tab:ARL1122}) for fixed values of $n=5$, $(\gamma_X,\gamma_Y)=(0.01,0.01)$, $\rho_0=\rho_1=-0.8$ and $\tau=0.99$, we
obtain $ARL_1=23.1$ when $(\Phi_{11},\Phi_{22})=(0.1,0.1)$ compared to $ARL_1=59.7$ when $(\Phi_{11},\Phi_{22})=(0.7,0.7)$. Both of these values are larger than the value 
$ARL_1=19.1$ when the process is free of
autocorrelation  (i.e.$(\Phi_{11},\Phi_{22})=(0,0)$) as pointed out in  \citet{Celano2014b} (Table 2). That is to say, the autocorrelation between observations has a negative influence on the Shewhart-RZ's performance: it redures the ability of the Shewhart-RZ control chart in detecting the process shift or out-of-control conditions.

 \begin{center}
INSERT TABLE \ref{tab:ARL1122} ABOUT HERE
\end{center}
 \begin{center}
INSERT TABLE \ref{tab:ARL1122r} ABOUT HERE
\end{center}
 \begin{center}
INSERT TABLE \ref{tab:ARL1122Phi} ABOUT HERE
\end{center}
 \begin{center}
INSERT TABLE \ref{tab:ARL1122rphi} ABOUT HERE
\end{center}

We also consider the situation when the occurrence of an assignable cause shifts the correlation coefficient $\rho$ from
$\rho_0$ to $\rho_1$. The values of  $ARL_1$ corresponding to this situation are presented in Tables~\ref{tab:ARL1221}-\ref{tab:ARL1221r}. For the brevity of the paper, we only show the $ARL_1$ values for the case $\Phi_{11}=\Phi_{22}$.  The negative effect of the autocorrelation on the chart's performance can also be seen from these two tables. Moreover, it is worth noting that when both the
nominal ratio and the correlation coefficient are shifted, for the process with high values of $(\gamma_X,\gamma_Y)$, the Shewhart-RZ control chart is in average inefficient in detecting shifts  as the values of $ARL_1$ are too large (see the case $(\gamma_X,\gamma_Y)=(0.2,0.2)$ in Table \ref{tab:ARL1221}).

 \begin{center}
INSERT TABLE \ref{tab:ARL1221} ABOUT HERE
\end{center}
 \begin{center}
INSERT TABLE \ref{tab:ARL1221r} ABOUT HERE
\end{center}


In several cases when there is no preference for any specific shift size, one may suggest assigning a distribution for the process shift in a predicted interval. The expected average run length ($EARL$) is used to evaluate the statistical performance of the corresponding chart, where
\begin{equation}
  \label{equ:EARL}
  EARL=\int_{\Omega}ARL\times f_{\tau}(\tau)\ud\tau,
\end{equation}
$f_{\tau}(\tau)$ stands for the ditribution of the shift size $\tau$ in the interal $\Omega$ and $ARL$ is as defined in \equ{equ:ARL}. When there is no information about $\tau$, a uniform distribution of $\tau$ could be applied, i.e. is the distribution of $f_{\tau}(\tau)=\frac{1}{b-a}$ for $\tau\in\Omega=[a,b]$.\\ 

\begin{figure}
\hspace*{-20mm}
  \includegraphics[width=170mm]{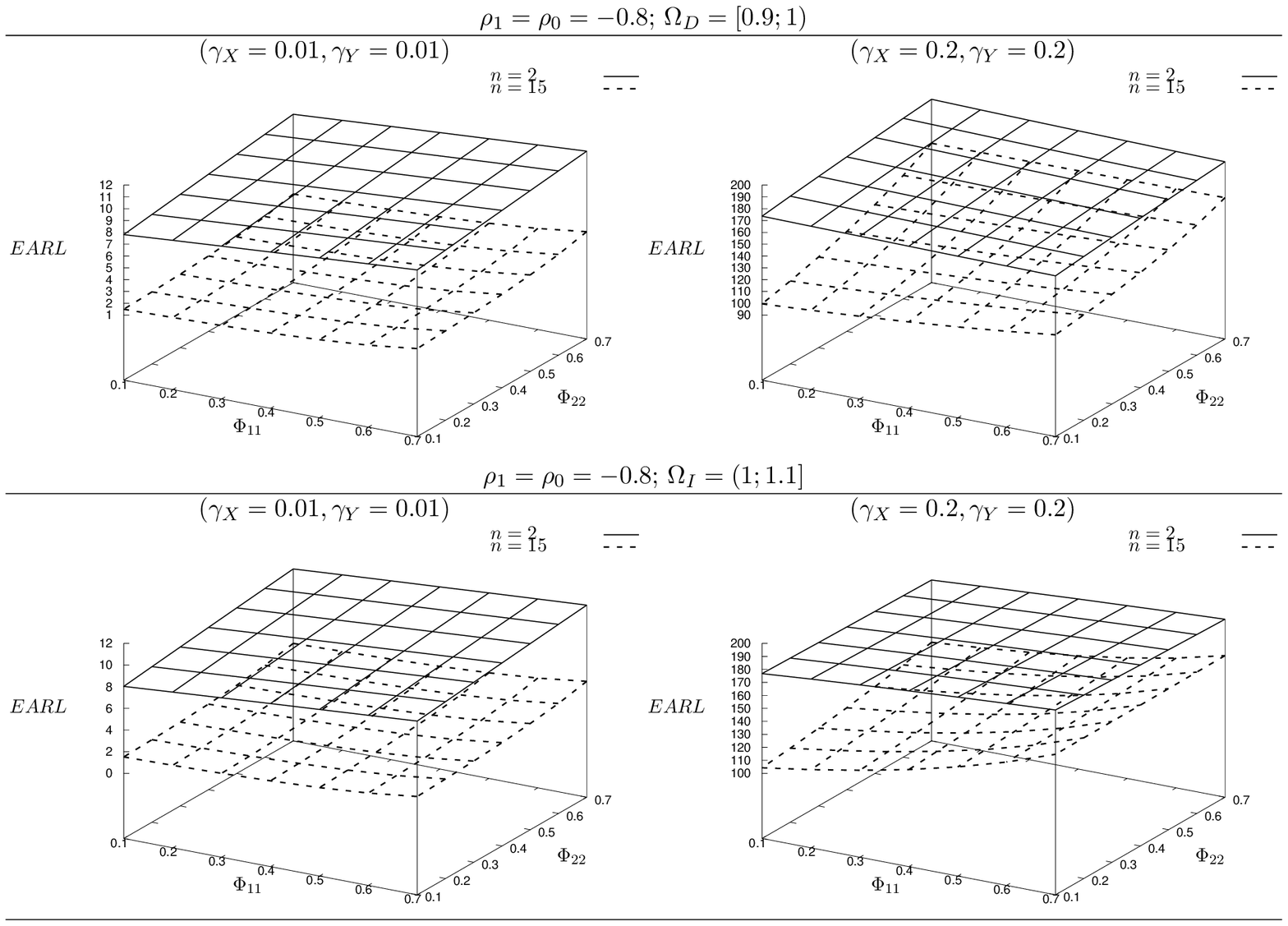} \\[5mm]
  \caption{The effects of $\Phi_{11}$ and $\Phi_{22}$ on the overall performance of the Shewhart-RZ control in the presence of autocorrelation
  chartr for  $n \in \{2,15\}$, $\gamma_X\in\{0.01,0.2\}$,
 $\gamma_Y\in\{0.01,0.2\}$, $\gamma_X=\gamma_Y$,  $\rho_0=\rho_1=-0.8$  and $ARL_0=200$.}
\label{fig:PHI}
\end{figure}

\begin{figure}
 \hspace*{-20mm}
    \includegraphics[width=170mm]{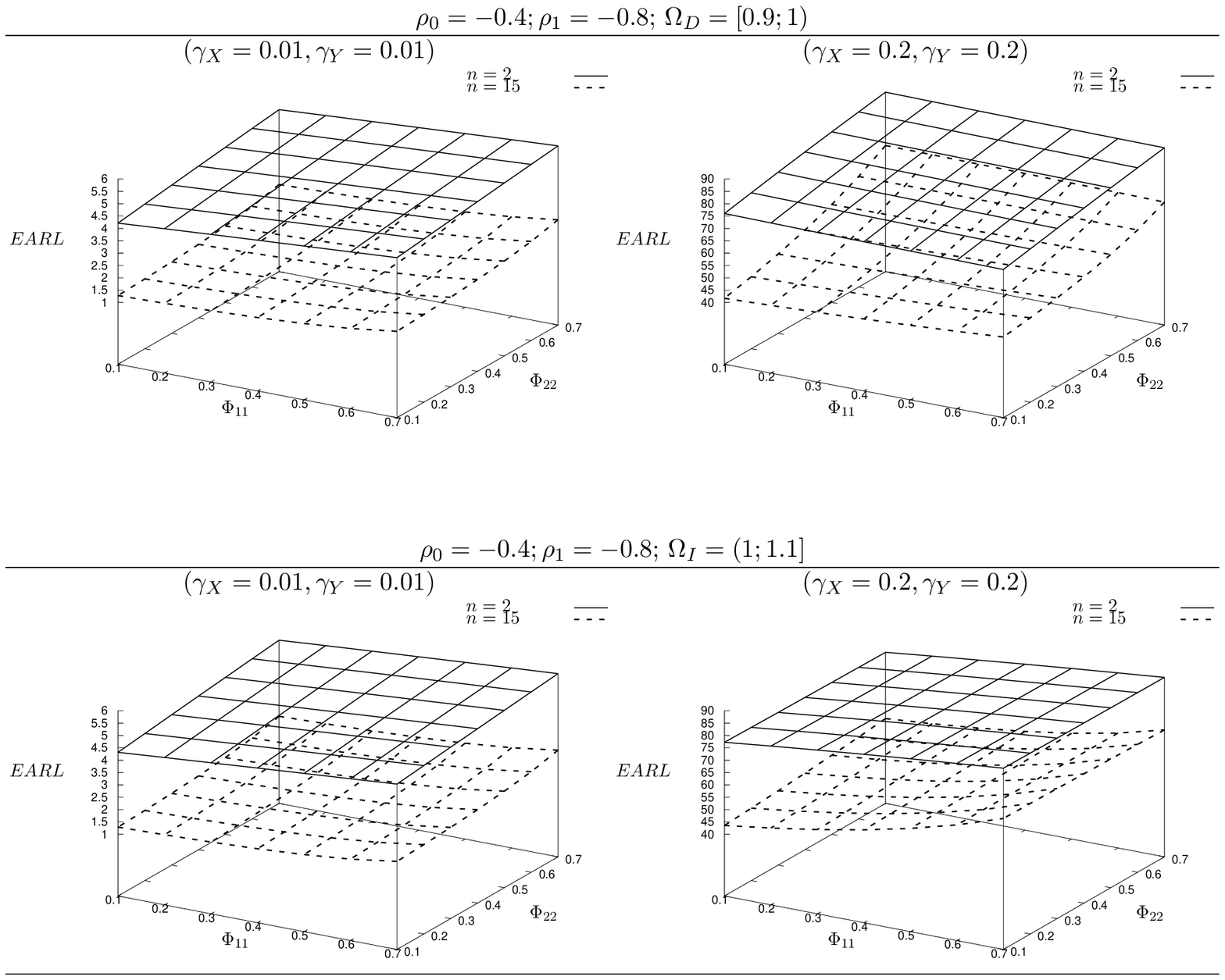} \\[5mm]
   \caption{The effects of $\Phi_{11}$ and $\Phi_{22}$ on the overall performance of the Shewhart-RZ control chart in the presence of autocorrelation
    for  $n\in\{2,15\}$, $\gamma_X\in\{0.01,0.2\}$,
 $\gamma_Y\in\{0.01,0.2\}$, $\gamma_X=\gamma_Y$, $\rho_0=-0.4$, $\rho_1=-0.8$ and $ARL_0=200$.}
\label{fig:PHIr}
\end{figure}

The effect of $\Phi_{11}$ and $\Phi_{22}$ on the overall performance of the Shewhart-RZ control chart in the presence of autocorrelation is displayed in Figure
\ref{fig:PHI} ($\rho_0=\rho_1=-0.8$) and Figure
\ref{fig:PHIr} ($\rho_0=-0.4$ and $\rho_1=-0.8$) with the $EARL$ values for two posibilities of $\Omega$, including $\Omega=\Omega_D=[0.9,1)$ (decreasing case, denoted 
as (D)) and $\Omega=\Omega_I=(1,1.1]$ (increasing case, denoted as (I) ). We also considre several values of  $\Phi_{11}$ and $\Phi_{22}$ as $\Phi_{11}\in\{0.1,0.2,\ldots,0.7\}$ and
$\Phi_{22}\in\{0.1,0.1,0.2,\ldots,0.7\}$. The values of others parameters are presented in the caption of each figure. In general, the obtained values of $EARL$ in these figures show a simimar trend of the negetive effect of $\Phi_{11}$ and $\Phi_{22}$ (representing the autocorrelation between observations) on the overall performance of Shewhart-RZ chart as the specifice shift size cases.
In most cases, the larger the values of ($\Phi_{11}$,$\Phi_{22}$), the larger the values of $EARL$, namely the faster the Shewhart-RZ
  control chart in detecting the of the out-of-control conditions. 
For example, in Figure \ref{fig:PHI} with $n=15$, $\Omega=[0.9,1)$, $\rho_0=\rho_1=-0.8$,
  $\gamma_X=\gamma_Y=0.01$, we have $EARL=1.49$ for
  $\Phi_{11}=\Phi_{22}=0.1$, $EARL=2.79$ for 
  $\Phi_{11}=\Phi_{22}=0.5$, and $EARL=4.57$ for 
  $\Phi_{11}=\Phi_{22}=0.7$. These values are all larger than $EARL=1.4$ for $\Phi_{11}=\Phi_{22}=$ (the process is free of
  autocorrelation) as presented in \citet{Celano2014b} (Table 6).
  

 \begin{center}
INSERT FIGURE \ref{fig:PHI} ABOUT HERE
\end{center}
 \begin{center}
INSERT FIGURE \ref{fig:PHIr} ABOUT HERE
\end{center}

\section{Illustrative example}
\label{sec:illustrative}
We consider a situation of monitoring the ratio of two normal variables in the food industry. In a mixture of various ingredients, pumpkin seeds and flaxseeds with added spices are two main components of a muesli brand recipe of a food company. Keeping equal weights of these two components in the recipe plays an important role in satisfying a requirement of food's nutrient composition. Also, the company has to produce brand boxes in different sizes to meet the market's needs. In the manufacturing process, the company would like to monitor on-line at regular intervals to check deviations from the in-control ratio for every size of boxes. This situation has been presented in \citet{Celano2014b} and then applied in a number of other studies such as \citet{Tran2015b}, \citet{Tran2016a}) and \citet{Nguyen_VSIEWMARZ_2018}.
 However, in these studies, the data was simulated based on an independent assumption between samples. In this section, the autocorrected data will be considered.   \\

 According to \citet{Celano2014b}, the in-control ratio $z_0=\frac{\mu_{p,i}}{\mu_{f,i}}=\frac{0.1}{0.1}=1$ is expected for all the box sizes, where $\mu_{p,i}$ and
$\mu_{f,i}$ stand for the mean weights of ``pumpkin seeds'' and ``flaxseeds''.  To monitor the ratio between these two components, suppose that a
sample of $n=5$ boxes is collected. 
The sample average weights
$\overline{W}_{p,i}=\frac{1}{n}\sum_{j=1}^nW_{p,i,j}$,  $\overline{W}_{f,i}=\frac{1}{n}\sum_{j=1}^nW_{f,i,j}$ and then the ratio 
$\hat{Z}_i=\frac{\overline{W}_{p,i}}{\overline{W}_{f,i}}$ are calculated where $X=W_{p,i,j}$ and $Y=W_{f,i,j}$ denote the weight of ``pumpkin seeds'' and ``flaxseeds''. The statistic $Z_i$ are plotted in the Shewhart-RZ chart to detect the out-of-control condition. We suppose that
\begin{itemize}
\item ($W_{p,i,j}, W_{f,i,j}$) can be well approximated by a bivariate normal variable. 
\item The observations are autocorrelated and follow the VAR(1) model with parameters of the VAR(1) model are 
\begin{equation}
\label{input}
\boldsymbol{\Phi}=\left(
  \begin{array}{cc}
    0.5 & 0 \\
    0 &  0.5
  \end{array}\right) ~ \text{and} ~~ \boldsymbol{\Sigma}_{\boldsymbol{\varepsilon}}=\left(
  \begin{array}{cc}
    0.0625 &  0.01\\ 
    0.01 & 0.0625
  \end{array}\right)
\end{equation} 
\end{itemize} 
In practice, these parameters of the VAR(1) model should be estimated from phase I with a sufficiently enough collected data. In this example, we suppose that the in-control ratio $z_0=1$ (i.e. $\mu_X=\mu_Y=25$ (gr)). Given the input parameters of the VAR(1) model in (\ref{input}), based on \eqref{equ:cova}, it is straightforward to obtain the standard deviation $\sigma_X$, $\sigma_Y$ of the weight of ``pumpkin seeds'' and ``flaxseeds''. Then, the parameters $\gamma_{\bar{X}}, \gamma_{\bar{Y}}, \bar{\rho}$ and $\bar{\omega}$ of the distribution of the sample average weights
$\overline{W}_{p,i}$ can be calculated from the formulas \equ{equ:gammabarX}--\equ{equ:omegabar}. 
 Eventually, from (\ref{equ:LCLRZ})-(\ref{equ:UCLRZ}), the control limits of the Shewhart-RZ chart in this case are calculated as $LCL=0.9723582$ and $UCL=1.0284276$. \\
 
\begin{figure}
  \caption{Shewhart-RZ control chart in the presence of autocorrelation for the food industry example}
  \begin{center}
    \includegraphics[width=120mm]{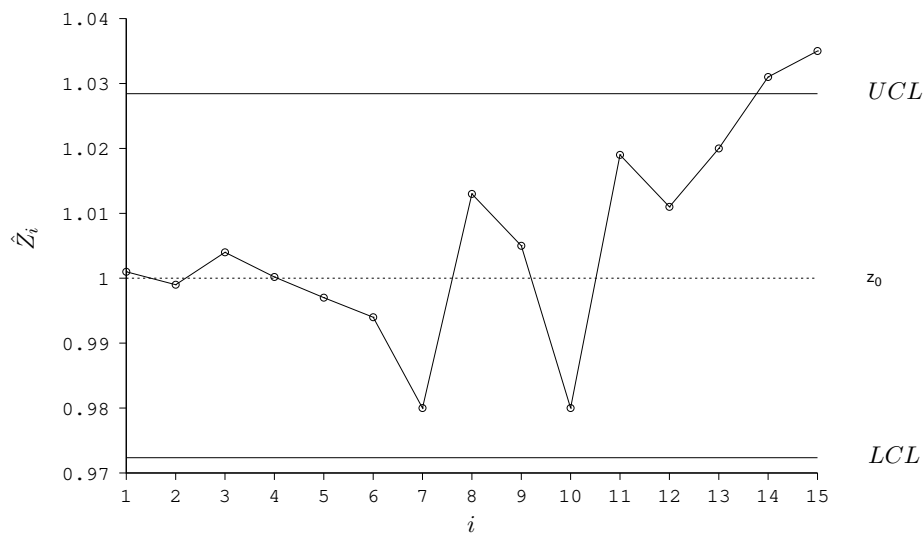} \\[5mm]
  \end{center}
\label{fig:mueslidata}
\end{figure}

The simulated data collected from the process are presented in Table \ref{tab:mueslidata}, where the values in the rightmost column are the statistics $\hat{Z}_i$. Similar to \cite{Celano2014b} we also assume that the process runs in-control condition  up to sample \#10 and  the
occurrence of an assignable cause is generated between samples \#10 and \#11 to shift the in-control ratio $z_0=1$ to $z_1 = 1.02$ (i.e ($\mu_X=25.05$ (gr), $\mu_Y=24.55$ (gr))), corresponding to a ratio percentage increase of 2\%.  The statistics $\hat{Z}_i$ are plotted on the Shewhart-RZ chart in \fig{fig:mueslidata}. As can be seen from this figure, the Shewhart-RZ chart signals the 
out-of-control conditions  by plotting samples \#14 and \#15 above the upper control limit. The Shewhart-RZ chart without considering the autocorrelation can detect the occurrence of the assignable cause from sample \#12 (corresponding to the upper control limit $UCL=0.018$).
 \begin{center}
INSERT TABLE \ref{tab:mueslidata} ABOUT HERE
\end{center}

\section{Conclusions}
\label{sec:conclusions}
In this paper, we have investigated the effects of autocorrelation on  the performance of Shewhart-RZ control chart using an 
autoregressive model for the sample ratio. Both  $ARL$ and
$EARL$ metrics are used to evaluate the performance of the Shewhart-RZ chart for the specific shift size (using $ARL$) or the overall performance (using $EARL$). The obtained results show that the autocorrelation between observations has a negative impact on the Shewhart-RZ chart's performance. It reduces the ability of the chart in detecting process shifts compared to no autocorrelation case. The presented results also show that the Shewhart control chart is not efficient in detecting small shifts for the process with high values of coefficient of variation (i.e. $\gamma_X,\gamma_Y)$). Designing other advanced control charts to reduce the negative impact of the autocorrelation on the chart's performance or investigating the effect of autocorrelation on the
performance of the EWMA-RZ control chart along similar lines, 
Phase I data implementation of RZ-type control charts could be interesting topics for further research.
\section*{Appendices}

\subsection*{Appendix 1}
Denote $\mathbf{W}=(X,Y)^T$ a bivariate normal random vector with mean vector $\boldsymbol{\mu_W}$ and variance-covariance matrix $\boldsymbol{\Sigma_W}$ where
\begin{equation}
  \label{equ:mean}
\boldsymbol{\mu_W}=\left(\begin{array}{c} \mu_X \\ \mu_Y \end{array}\right),
\end{equation}

\noindent
and 
\begin{equation}
  \label{equ:variance}
\boldsymbol{\Sigma_W}=\left(\begin{array}{cc}
\sigma_X^2 & \rho\sigma_X\sigma_Y \\
\rho\sigma_X\sigma_Y & \sigma_Y^2 \end{array} \right),
\end{equation}

The ratio of two components of $\mathbf{W}$, $X$ and $Y$, represents the ratio of two normal variables and is defined as $Z=\frac{X}{Y}$.  Several studies on the ditribution of this ratio has been published in the literature. In this study, we apply anapproximations for the inverse density function ($i.d.f$) $F^{-1}_Z(p|\gamma_X,\gamma_Y,\omega,\rho)$ of $Z$ proposed by \citet{Celano2014b} as:
\begin{equation}
  \label{equ:IDFZ}
  F^{-1}_Z(p|\gamma_X,\gamma_Y,\omega,\rho)\simeq\left\{
    \begin{array}{ll}
      \frac{-C_2-\sqrt{C_2^2-4C_1C_3}}{2C_1} & \mbox{if }p\in(0,0.5], \\
      \frac{-C_2+\sqrt{C_2^2-4C_1C_3}}{2C_1} & \mbox{if }p\in[0.5,1),
    \end{array}
  \right.
\end{equation}
where $C_1$, $C_2$ and $C_3$ are functions of $p$, $\gamma_X=\frac{\sigma_X}{\mu_X}$, $\gamma_Y=\frac{\sigma_Y}{\mu_Y}$,
$\omega=\frac{\sigma_X}{\sigma_Y}$ and $\rho$, i.e.
\begin{eqnarray}
C_1 & = & \frac{1}{\gamma_Y^2}-(\Phi^{-1}(p))^2, \\
C_2 & = & 2\omega\left(\rho(\Phi^{-1}(p))^2-\frac{1}{\gamma_X\gamma_Y}\right), \\
C_3 & = & \omega^2\left(\frac{1}{\gamma_X^2}-(\Phi^{-1}(p))^2\right),
\end{eqnarray}
In this paper, like in \citet{Celano2014b} we will assume that the coefficients of
variations $\gamma_X$ and $\gamma_Y$ are typically in the range
$(0,0.2]$. In this range, this approximation has been found to be accurate. 

{

\subsection*{Appendix 2}
To calculate $\mathrm{Vec}(\boldsymbol{\Sigma}_{\mathbf W})$ in  \eqref{equ:VecSigmaW}, we have,
\begin{align*}
\left(\mathbf{I_{4}}-\boldsymbol{\Phi}\otimes\boldsymbol{\Phi}\right)=\begin{pmatrix}
1-\Phi_{11}^2 & -\Phi_{11}\Phi_{12}& -\Phi_{12}\Phi_{11} & -\Phi_{12}^2\\
-\Phi_{11}\Phi_{21} & 1-\Phi_{11}\Phi_{22} & -\Phi_{12}\Phi_{21} & -\Phi_{12}\Phi_{22}\\
-\Phi_{21}\Phi_{11}& -\Phi_{21}\Phi_{12} & 1-\Phi_{22}\Phi_{11} & -\Phi_{22}\Phi_{12}\\
-\Phi_{21}^2 & -\Phi_{21}\Phi_{22} & -\Phi_{22}\Phi_{21} &1 -\Phi_{22}^2
\end{pmatrix}
\end{align*}

After some time-consuming calculations, one can  obtain:
\begin{align*}
\left(\mathbf{I_{4}}-\boldsymbol{\Phi}\otimes\boldsymbol{\Phi}\right)^{-1}=\frac{1}{\Delta}\begin{pmatrix}
\Delta_{11}  & \Delta_{12}& \Delta_{13} &\Delta_{14}\\
\Delta_{21} & \Delta_{22} & \Delta_{23} & \Delta_{24}\\
\Delta_{31}& \Delta_{32} & \Delta_{33}& \Delta_{34}\\
\Delta_{41} & \Delta_{42} & \Delta_{43}& \Delta_{44}
\end{pmatrix}
\end{align*}
where:
\begin{align*}
\Delta=(\Phi_{11}\Phi_{22}-\Phi_{12}\Phi_{21}-1)(\Phi_{11}^2\Phi_{22}^2-2\Phi_{11}\Phi_{12}\Phi_{21}\Phi_{22}+
\Phi_{12}^2\Phi_{21}^2-\Phi_{11}^2-2\Phi_{12}\Phi_{21}-\Phi_{22}^2+1)
\end{align*}
and 
\begin{align*}
\Delta_{11}&=-(\Phi_{11}\Phi_{22}^3-\Phi_{12}\Phi_{21}\Phi_{22}^2-\Phi_{11}\Phi_{22}-\Phi_{12}\Phi_{21}-\Phi_{22}^2+1)\\
\Delta_{12}&=\Delta_{13}=\Phi_{12}(\Phi_{11}\Phi_{22}^2-\Phi_{12}\Phi_{21}\Phi_{22}-\Phi_{11})\\
\Delta_{14}&=-\Phi_{12}^2(\Phi_{11}\Phi_{22}-\Phi_{12}\Phi_{21}+1)\\
\Delta_{21}&=\Phi_{21}(\Phi_{11}\Phi_{22}^2-\Phi_{12}\Phi_{21}\Phi_{22}-\Phi_{11})\\ \Delta_{22}&=-(\Phi_{11}^2\Phi_{22}^2-\Phi_{11}\Phi_{12}\Phi_{21}\Phi_{22}-\Phi_{11}^2-\Phi_{12}\Phi_{21}-\Phi_{22}^2+1)\\
\Delta_{23}&=-\Phi_{21}\Phi_{12}(\Phi_{11}\Phi_{22}-\Phi_{12}\Phi_{21}+1)\\
\Delta_{24}&=\Phi_{12}(\Phi_{11}^2\Phi_{22}-\Phi_{11}\Phi_{12}\Phi_{21}-\Phi_{22})\\
\Delta_{31}&=\Delta_{21}, \, \Delta_{32}=\Delta_{23},\, \Delta_{33}=\Delta_{22},\, \Delta_{34}=\Delta_{24}\\
\Delta_{41}&=-\Phi_{21}^2(\Phi_{11}\Phi_{22}-\Phi_{12}\Phi_{21}+1)\\
\Delta_{42}&=\Phi_{21}(\Phi_{11}^2\Phi_{22}-\Phi_{11}\Phi_{12}\Phi_{21}-\Phi_{22})\\
\Delta_{43}&=\Delta_{42}\\
\Delta_{44}&=-(\Phi_{11}^3\Phi_{22}-\Phi_{11}^2\Phi_{12}\Phi_{21}-\Phi_{11}^2-\Phi_{11}\Phi_{22}-\Phi_{12}\Phi_{21}+1) \\
\end{align*}

Accordingly, from these calculations and the formula of $\mathrm{Vec}(\boldsymbol{\Sigma}_{\mathbf W})$ in  \eqref{equ:VecSigmaW}, it is resulted that
\begin{align*}
\mathrm{Vec}(\boldsymbol{\Sigma}_{\mathbf W})&=\frac{1}{\Delta}\begin{pmatrix}
\Delta_{11}  & \Delta_{12}& \Delta_{13} &\Delta_{14}\\
\Delta_{21} & \Delta_{22} & \Delta_{23} & \Delta_{24}\\
\Delta_{31}& \Delta_{32} & \Delta_{33}& \Delta_{34}\\
\Delta_{41} & \Delta_{42} & \Delta_{43}& \Delta_{44}
\end{pmatrix} \begin{pmatrix}
 \sigma_{eX}^2\\
\sigma_{eXY} \\
 \sigma_{eXY}\\
 \sigma_{eY}^2
\end{pmatrix}\\
&=\frac{1}{\Delta}\begin{pmatrix}
\Delta_{11} \sigma_{eX}^2+ (\Delta_{12}+\Delta_{13}) \sigma_{eXY}+ \Delta_{14}   \sigma_{eY}^2\\
\Delta_{21} \sigma_{eX}^2+ (\Delta_{22}+\Delta_{23}) \sigma_{eXY}+ \Delta_{24}   \sigma_{eY}^2\\
\Delta_{31} \sigma_{eX}^2+ (\Delta_{32}+\Delta_{33}) \sigma_{eXY}+ \Delta_{34}   \sigma_{eY}^2\\
\Delta_{41} \sigma_{eX}^2+ (\Delta_{42}+\Delta_{43}) \sigma_{eXY}+ \Delta_{44}   \sigma_{eY}^2
\end{pmatrix}
\end{align*}

Finally, this leads us to obtain the variance-covariance matrix $\boldsymbol{\Sigma}_{\mathbf W}$ in  \eqref{equ:SigmaW}.

\subsection*{Appendix 3}
  Let   $\mathbf{W}_{i}=(W_{i,1},W_{i,2},....,W_{i,p})^{'}$  be  a stationary multivariate time series of dimension $p$  with mean vector  $\boldsymbol{\mu}_{\mathbf W}$ and  cross-covariances matrix  at lag $k$ as
\begin{align*} 
\Gamma(k)=E\big[(W_{i}-\boldsymbol{\mu}_{\mathbf W})(W_{i+k}-\boldsymbol{\mu}_{\mathbf W})^{\intercal}\big]_{p\times p}=\big[\gamma_{j,t}\big]_{j,t=1,2,...,p}
\end{align*}
where $\gamma_{jt}=E\left[(W_{i,j}-\mu_{j})(W_{i+k,t}-\mu_{t})\right]$ is the covariance between $W_{i,j}$ and $W_{i+k,t}$ for $k=0,\pm1,\pm2,\pm3,...$ . By definition, we have   $\boldsymbol{\Sigma}_{\mathbf{W}}=\Gamma(0)$. For the random sample $W_{i,1},W_{i,2},....,W_{i,n}$, \citet{Reinsel2003} showed that 
\begin{align*} 
\boldsymbol{\mu}_{\overline{\mathbf{W}}}&=\boldsymbol{\mu}_{\mathbf{W}}\\
\boldsymbol{\Sigma}_{\overline{\mathbf{W}}}&=\frac{1}{n^2}\sum_{j=1}^{n}\sum_{t=1}^{n}\Gamma(j-t)=\frac{1}{n^2}\sum_{k=-(n-1)}^{n-1}\left(n-|k|\right)\Gamma(k).
\end{align*}

 Adopting  this result for   VAR(1) autoregressive   model, $\pmb{\Sigma}_{\overline{\textbf{W}}}$   can be obtained as
\begin{small}
\begin{align}\label{equ:A3VCwbar}
\boldsymbol{\Sigma}_{\overline{\mathbf{W}}}&=\frac{1}{n}\boldsymbol{\Sigma}_{\mathbf{W}}+\frac{1}{n^2}\sum_{k=1}^{n-1}(n-k)\Gamma(k)+\frac{1}{n^2}\sum_{k=1}^{n-1}(n-k)\Gamma(-k)  \nonumber\\
&=\frac{1}{n}\boldsymbol{\Sigma}_{\mathbf{W}}+\frac{1}{n^2}\sum_{k=1}^{n-1}(n-k)[\Gamma(k)+\Gamma(k)^{\intercal}] \qquad \qquad \left(\Gamma(-k)=\Gamma(k)^{\intercal}\right) \nonumber\\
&=\frac{1}{n}\boldsymbol{\Sigma}_{\mathbf W}+\frac{1}{n^2} \sum_{k=1}^{n-1}(n-k) \Big[\boldsymbol{\Sigma}_{\mathbf W}\boldsymbol{\Phi}^{\intercal^{k}}+\boldsymbol{\Phi}^k \boldsymbol{\Sigma}_{\mathbf{W}}^{\intercal} \Big] 
\qquad \qquad \left(\Gamma(k)=\boldsymbol{\Sigma}_{\mathbf{W}}\boldsymbol{\Phi}^{\intercal^k}\right) \nonumber\\
&=\frac{1}{n}\boldsymbol{\Sigma}_{\mathbf{W}}+\frac{1}{n}\sum_{k=1}^{n-1}\Big[\boldsymbol{\Sigma}_{\mathbf{W}}\boldsymbol{\Phi}^{\intercal^{k}}+\boldsymbol{\Phi}^k \boldsymbol{\Sigma}_{\mathbf{W}}^{\intercal}\Big]-\frac{1}{n^2} \sum_{k=1}^{n-1} k\Big[\boldsymbol{\Sigma}_{\mathbf{W}}\boldsymbol{\Phi}^{\intercal^{k}}+\boldsymbol{\Phi}^k \boldsymbol{\Sigma}_{\mathbf{W}}^{\intercal}\Big].
\end{align} 
\end{small}

In order to further simplify the above relation, we use proposition 1.5.38. of \citet{Hubbard2015} in which they  proved  that   $\sum_{n=0}^{\infty} \boldsymbol{\Phi}^{n}=(\mathbf{I}_{p}-\boldsymbol{\Phi})^{-1}$ for     a $p \times p$ square matrix  $\boldsymbol{\Phi}$    that   the absolute values of all its eigenvalues are  less than 1. Here, we know that this condition about  $\boldsymbol{\Phi}$ is hold because it is   the necessary and sufficient condition for the stationarity of a VAR(1) autoregressive model.  Using the idea of their proof, it can be showed that $\sum_{k=1}^{n-1} \boldsymbol{\Phi}^{k}=\Lambda(\boldsymbol{\Phi})$ and $\sum_{k=1}^{n-1} k\boldsymbol{\Phi}^{k}=\Pi(\boldsymbol{\Phi})$ where 
\begin{align*} 
\Lambda(\boldsymbol{\Phi})&=(\boldsymbol{\Phi}-\boldsymbol{\Phi}^{n})(\mathbf{I}_{2}-\boldsymbol{\Phi})^{-1}\\\label{equ:ssum}
\Pi(\boldsymbol{\Phi})&=(\boldsymbol{\Phi}^{-1}-\mathbf{I}_{2})^{-1}\Big((\mathbf{I}_{2}-\boldsymbol{\Phi}^{n-1})(\mathbf{I}_{2}-\boldsymbol{\Phi})^{-1}-(n-1)\boldsymbol{\Phi}^{n-1}\Big)
\end{align*}
 
Rewriting \eqref{equ:A3VCwbar} based on $\Lambda(\boldsymbol{\Phi})$ and $\Pi(\boldsymbol{\Phi})$ and some mathematical calculations result $\boldsymbol{\Sigma}_{\overline{\mathbf{W}}}$ in \eqref{equ:VCwbar}.

 }

\begin{table}

 \caption{Values of $LCL$ (first row) and $UCL$ (second row) for the Shewhart-RZ control chart 
  in the presence of autocorrelation, for $z_0=1$, $ARL_0=200$, $\Phi_{11}=\Phi_{22}=0.1$, 
  $n\in\{2,5,7,10,15\}$, $\gamma_X\in\{0.01,0.2\}$,
  $\gamma_Y\in\{0.01,0.2\}$ and $\rho_0\in\{-0.8,-0.4,0,0.4,0.8\}$.}
  \hspace{10mm}
   \scalebox{0.95}{
\begin{tabular}{cccccccc}
      \hline
      $\gamma_X$ & $\gamma_Y$ & $\rho_0$ & $n=1$ & $n=5$ & $n=7$ & $n=10$& $n=15$ \\
      \hline
$0.01$ & $0.01$ & $-0.8$ & $0.9596$ & $0.9725$ & $0.9764$ & $0.9800$ & $0.9835$ \\
       &        &        & $1.0421$ & $1.0283$ & $1.0242$ & $1.0204$ & $1.0167$ \\
 &  & $-0.4$ & $0.9643$ & $0.9757$ & $0.9791$ & $0.9823$ & $0.9855$ \\
       &        &        & $1.0371$ & $1.0249$ & $1.0213$ & $1.0180$ & $1.0148$ \\
 &  & $ 0.0$ & $0.9697$ & $0.9794$ & $0.9823$ & $0.9851$ & $0.9877$ \\
       &        &        & $1.0312$ & $1.0210$ & $1.0180$ & $1.0152$ & $1.0125$ \\
 &  & $ 0.4$ & $0.9765$ & $0.9840$ & $0.9863$ & $0.9884$ & $0.9905$ \\
\hline
  \end{tabular}}
\label{tab:LUCL}
\end{table}

\begin{sidewaystable}
  \hspace*{5mm}
  \scalebox{0.7}
  {\begin{tabular}{ccccccccccccc}
      \hline
      & & \multicolumn{5}{c}{$(\gamma_X=0.01,\gamma_Y=0.01)$} && \multicolumn{5}{c}{$(\gamma_X=0.2,\gamma_Y=0.2)$} \\
        $  $ & & $n=2$ & $n=5$ & $n=7$ & $n=10$ & $n=15$ & & $n=2$ & $n=5$ & $n=7$ & $n=10$ & $n=15$ \\
     \cline{1-1}\cline{3-7}\cline{9-13}
\cline{1-1}\cline{3-7}\cline{9-13}
& & \multicolumn{11}{c}{$\rho_0=\rho_1=-0.8$} \\
$0.90$ &  & ($  1.0$,$  1.0$) & ($  1.0$,$  1.0$) & ($  1.0$,$  1.0$) & ($  1.0$,$  1.0$) & ($  1.0$,$  1.0$) &  & ($137
.4$,$162.2$) & ($ 83.1$,$144.2$) & ($ 64.2$,$133.7$) & ($ 46.7$,$119.6$) & ($ 30.7$,$100.5$) \\
$0.95$ &  & ($  1.3$,$  1.8$) & ($  1.0$,$  1.4$) & ($  1.0$,$  1.2$) & ($  1.0$,$  1.1$) & ($  1.0$,$  1.0$) &  & ($181
.2$,$189.8$) & ($153.4$,$183.7$) & ($138.6$,$179.7$) & ($120.7$,$173.8$) & ($ 98.5$,$164.1$) \\
$0.98$ &  & ($ 11.8$,$ 20.3$) & ($  3.7$,$ 13.6$) & ($  2.4$,$ 10.9$) & ($  1.6$,$  8.1$) & ($  1.2$,$  5.4$) &  & ($196
.9$,$198.4$) & ($191.2$,$197.3$) & ($187.6$,$196.6$) & ($182.3$,$195.5$) & ($174.1$,$193.6$) \\
$0.99$ &  & ($ 54.4$,$ 76.1$) & ($ 23.1$,$ 59.7$) & ($ 15.6$,$ 51.7$) & ($  9.9$,$ 42.2$) & ($  5.7$,$ 31.3$) &  & ($199
.2$,$199.6$) & ($197.8$,$199.3$) & ($196.8$,$199.1$) & ($195.3$,$198.9$) & ($193.0$,$198.4$) \\
$1.01$ &  & ($ 55.3$,$ 77.2$) & ($ 23.6$,$ 60.7$) & ($ 16.0$,$ 52.6$) & ($ 10.2$,$ 43.0$) & ($  5.9$,$ 32.0$) &  & ($199
.2$,$199.6$) & ($197.8$,$199.3$) & ($196.8$,$199.2$) & ($195.4$,$198.9$) & ($193.1$,$198.4$) \\
\hline
    \end{tabular} }
      \caption{The $ARL_1$ values of the Shewhart-RZ in the presence of autocorrelation for
      $\Phi_{11}=\Phi_{22}=0.1$ (left side),  $\Phi_{11}=\Phi_{22}=0.7$ (right side), $\gamma_X= \gamma_Y \in\{0.01,0.2\}$, $\rho_0\in\{-0.8,-0.4,0,0.4,0.8\}$,
    $\rho_0=\rho_1$,
    $n\in\{2,5,7,10,15\}$ and $ARL_0=200$.}
  \label{tab:ARL1122}
\end{sidewaystable}

\begin{sidewaystable}
\hspace*{5mm}
  \scalebox{0.7}
  {\begin{tabular}{ccccccccccccc}
      \hline
     & & \multicolumn{5}{c}{$(\gamma_X=0.01,\gamma_Y=0.2)$} && \multicolumn{5}{c}{$(\gamma_X=0.2,\gamma_Y=0.01)$} \\
        $  $ & & $n=1$ & $n=5$ & $n=7$ & $n=10$ & $n=15$ & & $n=1$ & $n=5$ & $n=7$ & $n=10$ & $n=15$ \\
\cline{1-1}\cline{3-7}\cline{9-13}
& & \multicolumn{11}{c}{$\rho_0=\rho_1=-0.8$} \\
$0.90$ &  & ($ 34.0$,$ 44.7$) & ($ 16.6$,$ 36.7$) & ($ 11.9$,$ 32.6$) & ($  8.0$,$ 27.5$) & ($  5.0$,$ 21.4$) &  & ($114
.8$,$167.9$) & ($ 43.7$,$127.6$) & ($ 28.1$,$108.3$) & ($ 16.8$,$ 86.3$) & ($  9.0$,$ 61.7$) \\
$0.95$ &  & ($ 94.2$,$105.8$) & ($ 65.2$,$ 97.5$) & ($ 53.5$,$ 92.4$) & ($ 41.5$,$ 85.4$) & ($ 29.5$,$ 75.2$) &  & ($195
.5$,$225.9$) & ($125.6$,$203.9$) & ($ 99.4$,$191.0$) & ($ 74.0$,$173.4$) & ($ 49.8$,$148.7$) \\
$0.98$ &  & ($159.3$,$163.2$) & ($146.0$,$160.5$) & ($138.2$,$158.7$) & ($127.8$,$155.9$) & ($113.3$,$151.3$) &  & ($218
.8$,$224.6$) & ($198.8$,$220.5$) & ($187.1$,$217.8$) & ($171.8$,$213.6$) & ($150.7$,$206.7$) \\
$0.99$ &  & ($181.4$,$182.5$) & ($177.2$,$181.7$) & ($174.4$,$181.2$) & ($170.5$,$180.4$) & ($164.2$,$178.9$) &  & ($213
.0$,$214.4$) & ($207.7$,$213.4$) & ($204.4$,$212.7$) & ($199.5$,$211.7$) & ($191.8$,$209.9$) \\
$1.01$ &  & ($212.9$,$214.2$) & ($207.7$,$213.3$) & ($204.4$,$212.6$) & ($199.6$,$211.6$) & ($192.1$,$209.9$) &  & ($181
.6$,$182.7$) & ($177.4$,$181.9$) & ($174.7$,$181.4$) & ($170.9$,$180.6$) & ($164.7$,$179.2$) \\
\hline
    \end{tabular} }
  \caption{The $ARL_1$ values of the Shewhart-RZ in the presence of autocorrelation for
      $\Phi_{11}=\Phi_{22}=0.1$ (left side),  $\Phi_{11}=\Phi_{22}=0.7$ (right side), $\gamma_X\in\{0.01,0.2\}$,
 $\gamma_Y\in\{0.01,0.2\}$, $\gamma_X\neq \gamma_Y$, $\rho_0\in\{-0.8,-0.4,0,0.4,0.8\}$,
    $\rho_0=\rho_1$,
    $n\in\{2,5,7,10,15\}$ and $ARL_0=200$.}
  \label{tab:ARL1122r}
\end{sidewaystable}

\begin{sidewaystable}
  \hspace*{5mm}
  \scalebox{0.7}
  {\begin{tabular}{ccccccccccccc}
      \hline
      & & \multicolumn{5}{c}{$(\gamma_X=0.01,\gamma_Y=0.01)$} && \multicolumn{5}{c}{$(\gamma_X=0.2,\gamma_Y=0.2)$} \\
        $  $ & & $n=2$ & $n=5$ & $n=7$ & $n=10$ & $n=15$ & & $n=2$ & $n=5$ & $n=7$ & $n=10$ & $n=15$ \\
     \cline{1-1}\cline{3-7}\cline{9-13}
\cline{1-1}\cline{3-7}\cline{9-13}
& & \multicolumn{11}{c}{$\rho_0=\rho_1=-0.8$} \\
\cline{1-1}\cline{3-7}\cline{9-13}
& & \multicolumn{11}{c}{$\rho_0=\rho_1=-0.8$} \\
$0.90$ &  & ($  1.0$,$  1.0$) & ($  1.0$,$  1.0$) & ($  1.0$,$  1.0$) & ($  1.0$,$  1.0$) & ($  1.0$,$  1.0$) &  & ($136
.9$,$167.2$) & ($ 98.4$,$146.2$) & ($ 85.5$,$131.9$) & ($ 71.9$,$113.2$) & ($ 56.8$,$ 89.3$) \\
$0.95$ &  & ($  1.5$,$  1.5$) & ($  1.1$,$  1.1$) & ($  1.0$,$  1.0$) & ($  1.0$,$  1.0$) & ($  1.0$,$  1.0$) &  & ($176
.9$,$196.0$) & ($156.3$,$193.3$) & ($148.3$,$188.2$) & ($139.0$,$179.4$) & ($126.5$,$164.8$) \\
$0.98$ &  & ($ 15.9$,$ 16.2$) & ($  8.2$,$  8.5$) & ($  6.1$,$  6.3$) & ($  4.3$,$  4.5$) & ($  2.9$,$  2.9$) &  & ($193
.7$,$201.8$) & ($187.2$,$204.1$) & ($184.9$,$203.9$) & ($182.2$,$202.7$) & ($178.7$,$200.0$) \\
$0.99$ &  & ($ 65.6$,$ 66.6$) & ($ 42.5$,$ 43.8$) & ($ 34.5$,$ 35.6$) & ($ 26.4$,$ 27.2$) & ($ 18.2$,$ 18.8$) &  & ($197
.4$,$201.5$) & ($194.6$,$203.2$) & ($193.7$,$203.5$) & ($192.8$,$203.4$) & ($191.7$,$202.9$) \\
\hline
    \end{tabular} }
      \caption{$ARL_1$ values of the Shewhart-RZ in the presence of autocorrelation  for
      $\Phi_{11}=0.1,\Phi_{22}=0.7$ (left side), $\Phi_{11}=0.7, \Phi_{22}=0.1$ (right side), $\gamma_X\in\{0.01,0.2\}$,
 $\gamma_Y\in\{0.01,0.2\}$, $\gamma_X=\gamma_Y$, $\rho_0\in\{-0.8,-0.4,0,0.4,0.8\}$,
    $\rho_0=\rho_1$,
    $n\in\{2,5,7,10,15\}$ and $ARL_0=200$.}
  \label{tab:ARL1122Phi}
\end{sidewaystable}

\begin{sidewaystable}
\hspace*{5mm}
  \scalebox{0.7}
  {\begin{tabular}{ccccccccccccc}
      \hline
     & & \multicolumn{5}{c}{$(\gamma_X=0.01,\gamma_Y=0.2)$} && \multicolumn{5}{c}{$(\gamma_X=0.2,\gamma_Y=0.01)$} \\
        $  $ & & $n=1$ & $n=5$ & $n=7$ & $n=10$ & $n=15$ & & $n=1$ & $n=5$ & $n=7$ & $n=10$ & $n=15$ \\
\cline{1-1}\cline{3-7}\cline{9-13}
& & \multicolumn{11}{c}{$\rho_0=\rho_1=-0.8$} \\
$0.90$ &  & ($ 44.0$,$ 35.0$) & ($ 35.5$,$ 18.1$) & ($ 31.4$,$ 13.3$) & ($ 26.5$,$  9.2$) & ($ 20.6$,$  5.9$) &  & ($115
.6$,$167.6$) & ($ 45.8$,$126.7$) & ($ 30.1$,$107.3$) & ($ 18.5$,$ 85.2$) & ($ 10.2$,$ 60.6$) \\
$0.95$ &  & ($104.9$,$ 95.6$) & ($ 95.8$,$ 69.0$) & ($ 90.6$,$ 57.7$) & ($ 83.5$,$ 45.8$) & ($ 73.3$,$ 33.4$) &  & ($195
.6$,$226.2$) & ($128.1$,$204.0$) & ($102.9$,$190.7$) & ($ 78.2$,$172.8$) & ($ 54.0$,$147.8$) \\
$0.98$ &  & ($162.7$,$160.1$) & ($159.5$,$148.8$) & ($157.6$,$142.0$) & ($154.8$,$132.7$) & ($150.1$,$119.3$) &  & ($218
.3$,$225.0$) & ($198.8$,$221.0$) & ($187.9$,$218.3$) & ($173.8$,$214.0$) & ($154.3$,$206.9$) \\
$0.99$ &  & ($182.3$,$181.8$) & ($181.3$,$178.5$) & ($180.7$,$176.3$) & ($179.8$,$172.9$) & ($178.3$,$167.5$) &  & ($212
.7$,$214.6$) & ($207.2$,$213.7$) & ($204.0$,$213.1$) & ($199.5$,$212.0$) & ($192.5$,$210.2$) \\
$1.01$ &  & ($214.5$,$212.6$) & ($213.6$,$207.2$) & ($213.0$,$204.1$) & ($212.0$,$199.6$) & ($210.2$,$192.8$) &  & ($182
.0$,$182.5$) & ($178.8$,$181.5$) & ($176.6$,$180.9$) & ($173.3$,$180.0$) & ($167.9$,$178.5$) \\
\hline
    \end{tabular} }
  \caption{$ARL_1$ values of the Shewhart-RZ in the presence of autocorrelation for
      $\Phi_{11}=0.1, \Phi_{22}=0.7$ (left side), $\Phi_{11}=0.7, \Phi_{22}=0.1$ (right side),  $\gamma_X\in\{0.01,0.2\}$,
 $\gamma_Y\in\{0.01,0.2\}$, $\gamma_X \neq \gamma_Y$, $\rho_0\in\{-0.8,-0.4,0,0.4,0.8\}$,
    $\rho_0=\rho_1$,
    $n\in\{2,5,7,10,15\}$ and $ARL_0=200$.}
  \label{tab:ARL1122rphi}
\end{sidewaystable}

\begin{sidewaystable}
  \scalebox{0.75}
  {\begin{tabular}{ccccccccccccc}
      \hline
       & & \multicolumn{5}{c}{$(\gamma_X=0.01,\gamma_Y=0.01)$} && \multicolumn{5}{c}{$(\gamma_X=0.2,\gamma_Y=0.2)$} \\
        $  $ & & $n=2$ & $n=5$ & $n=7$ & $n=10$ & $n=15$ & & $n=2$ & $n=5$ & $n=7$ & $n=10$ & $n=15$ \\
\cline{1-1}\cline{3-7}\cline{9-13}
& & \multicolumn{11}{c}{$\rho_0=-0.4,\rho_1=-0.2$} \\
$0.90$ &  & ($  1.0$,$  1.0$) & ($  1.0$,$  1.0$) & ($  1.0$,$  1.0$) & ($  1.0$,$  1.0$) & ($  1.0$,$  1.0$) &  & ($209
.0$,$249.2$) & ($110.2$,$221.2$) & ($ 79.0$,$202.3$) & ($ 52.8$,$176.3$) & ($ 31.4$,$140.9$) \\
$0.95$ &  & ($  1.1$,$  1.4$) & ($  1.0$,$  1.1$) & ($  1.0$,$  1.1$) & ($  1.0$,$  1.0$) & ($  1.0$,$  1.0$) &  & ($308
.7$,$308.6$) & ($256.9$,$310.6$) & ($223.9$,$307.2$) & ($184.8$,$298.6$) & ($140.0$,$280.2$) \\
$0.98$ &  & ($  9.9$,$ 18.9$) & ($  2.7$,$ 11.7$) & ($  1.8$,$  9.0$) & ($  1.3$,$  6.4$) & ($  1.1$,$  4.1$) &  & ($349
.0$,$328.1$) & ($360.1$,$344.6$) & ($354.6$,$351.0$) & ($342.6$,$356.8$) & ($321.1$,$360.6$) \\
$0.99$ &  & ($ 64.4$,$ 99.7$) & ($ 22.1$,$ 72.6$) & ($ 13.8$,$ 60.3$) & ($  8.1$,$ 46.7$) & ($  4.4$,$ 32.2$) &  & ($355
.3$,$331.0$) & ($380.1$,$349.8$) & ($383.3$,$357.9$) & ($383.5$,$366.4$) & ($379.5$,$374.9$) \\
$1.01$ &  & ($ 65.8$,$101.5$) & ($ 22.7$,$ 74.2$) & ($ 14.2$,$ 61.6$) & ($  8.3$,$ 47.8$) & ($  4.5$,$ 33.0$) &  & ($355
.3$,$331.0$) & ($380.2$,$349.9$) & ($383.5$,$357.9$) & ($383.8$,$366.5$) & ($379.9$,$375.0$) \\
\hline
    \end{tabular} }
    \caption{$ARL_1$ values of the Shewhart-RZ in the presence of autocorrelation for
      $\Phi_{11}=\Phi_{22}=0.1$ (left side), $\Phi_{11}=\Phi_{22}=0.7$ (right side), $\gamma_X\in\{0.01,0.2\}$,
 $\gamma_X\in\{0.01,0.2\}$, $\gamma_X=\gamma_Y$, $(\rho_0,\rho_1)=\{(-0.4,-0.2),(-0.4,-0.8),(0.4,0.2),(0.4,0.8)\}$,
    $n\in\{2,5,7,10,15\}$ and $ARL_0=200$.}
  \label{tab:ARL1221}
\end{sidewaystable}

\begin{sidewaystable}
  \scalebox{0.7}
  {\begin{tabular}{ccccccccccccc}
      \hline
 & & \multicolumn{5}{c}{$(\gamma_X=0.01,\gamma_Y=0.2)$} && \multicolumn{5}{c}{$(\gamma_X=0.2,\gamma_Y=0.01)$} \\
        $  $ & & $n=2$ & $n=5$ & $n=7$ & $n=10$ & $n=15$ & & $n=2$ & $n=5$ & $n=7$ & $n=10$ & $n=15$ \\
\cline{1-1}\cline{3-7}\cline{9-13}
& & \multicolumn{11}{c}{$\rho_0=-0.4,\rho_1=-0.2$} \\
$0.90$ &  & ($ 34.2$,$ 45.5$) & ($ 16.2$,$ 37.0$) & ($ 11.5$,$ 32.7$) & ($  7.8$,$ 27.4$) & ($  4.8$,$ 21.2$) &  & ($118
.1$,$173.9$) & ($ 44.0$,$131.6$) & ($ 28.0$,$111.4$) & ($ 16.5$,$ 88.3$) & ($  8.7$,$ 62.7$) \\
$0.95$ &  & ($ 99.3$,$112.5$) & ($ 67.3$,$103.0$) & ($ 54.7$,$ 97.3$) & ($ 42.0$,$ 89.4$) & ($ 29.4$,$ 78.2$) &  & ($206
.7$,$240.0$) & ($130.8$,$215.9$) & ($102.7$,$201.7$) & ($ 75.8$,$182.5$) & ($ 50.4$,$155.7$) \\
$0.98$ &  & ($172.4$,$176.8$) & ($157.0$,$173.7$) & ($148.0$,$171.6$) & ($136.2$,$168.4$) & ($119.8$,$163.1$) &  & ($236
.4$,$243.0$) & ($213.6$,$238.4$) & ($200.4$,$235.2$) & ($183.1$,$230.5$) & ($159.5$,$222.7$) \\
$0.99$ &  & ($197.1$,$198.4$) & ($192.2$,$197.5$) & ($189.1$,$196.9$) & ($184.5$,$196.0$) & ($177.2$,$194.3$) &  & ($231
.2$,$232.8$) & ($225.3$,$231.7$) & ($221.4$,$231.0$) & ($215.7$,$229.8$) & ($206.9$,$227.8$) \\
$1.01$ &  & ($231.2$,$232.7$) & ($225.3$,$231.6$) & ($221.5$,$230.9$) & ($215.9$,$229.7$) & ($207.3$,$227.7$) &  & ($197
.4$,$198.6$) & ($192.6$,$197.7$) & ($189.4$,$197.1$) & ($184.9$,$196.2$) & ($177.8$,$194.6$) \\
\hline
    \end{tabular} }
        \caption{$ARL_1$ values of the Shewhart-RZ in the presence of autocorrelation  for
      $\Phi_{11}=\Phi_{22}=0.1$ (left side),  $\Phi_{11}=\Phi_{22}=0.7$ (right side), $\gamma_X\in\{0.01,0.2\}$,
 $\gamma_X\in\{0.01,0.2\}$, $\gamma_X \neq \gamma_Y$, $(\rho_0,\rho_1)=\{(-0.4,-0.2),(-0.4,-0.8),(0.4,0.2),(0.4,0.8)\}$,
    $n\in\{2,5,7,10,15\}$ and $ARL_0=200$.}
  \label{tab:ARL1221r}
\end{sidewaystable}

\begin{table}
  \begin{center}
\scalebox{0.8}{
  \begin{tabular}{c|ccccc|cc}
    \hline
    \multirow{2}{*}{Sample}&  \multicolumn{5}{c|}{$W_{p,i,j}$ [gr]} & $\bar{W}_{p,i}$ [gr] & \multirow{2}{*}{ $\hat{Z_i}=\frac{\bar{W}_{p,i}}{\bar{W}_{f,i}}$} \\
    & \multicolumn{5}{c|}{$W_{f,i,j}$ [gr]} & $\bar{W}_{f,i}$ [gr] &\\
    \hline
    \multirow{2}{*}{1} &    24.923& 24.784& 24.756 &24.576& 25.154&24.839& \multirow{2}{*}{1.001} \\
&25.086& 25.061& 24.645& 24.419 &24.&24.814\\[2mm]
    \multirow{2}{*}{2}   & 24.735& 24.973& 25.167& 25.084& 24.710&24.934& \multirow{2}{*}{0.999}\\
&24.463& 25.489& 25.355& 24.725& 24.676&24.942\\[2mm]

    \multirow{2}{*}{3} &    25.071& 24.600& 25.008& 25.129& 25.364& 25.034& \multirow{2}{*}{1.004}\\
& 24.455& 24.684& 25.240& 25.170& 25.033& 24.916\\[2mm]  
    \multirow{2}{*}{4}&   25.591& 25.122& 25.006& 24.772& 24.504& 24.999& \multirow{2}{*}{1.0002}\\
& 25.440& 25.057& 24.878& 24.762& 24.832& 24.994\\[2mm]
\multirow{2}{*}{5}   &  24.695& 25.094& 24.787& 24.941& 25.261& 24.95& \multirow{2}{*}{0.997}\\
 &24.942& 25.347& 25.001&24.646& 25.178& 25.023\\[2mm]
    \multirow{2}{*}{6}  &   25.087& 25.080& 25.366& 25.225& 25.118&25.175&\multirow{2}{*}{0.994}\\
&25.219& 25.433& 25.431& 25.057& 25.478&  25.324\\[2mm]
    \multirow{2}{*}{7}  &  24.675& 25.011&24.809& 24.919& 24.740& 24.831& \multirow{2}{*}{0.980}\\
&25.169& 25.749& 24.930& 25.234& 25.506& 25.318\\[2mm]
    \multirow{2}{*}{8}& 25.114& 25.379&v25.491& 25.376& 25.050& 25.2& \multirow{2}{*}{1.013}\\
& 24.889& 24.822& 25.123& 24.785& 25.072& 24.93 \\[2mm]  
    \multirow{2}{*}{9}  & 25.223& 25.036& 24.880& 24.992& 25.055& 25.037& \multirow{2}{*}{1.005}\\
&25.001& 24.923& 24.964& 24.913& 24.729& 24.906 \\[2mm]
    \multirow{2}{*}{10} &  24.955& 24.985& 24.720& 24.488& 24.943&  24.818& \multirow{2}{*}{0.980}\\
&25.555& 25.458& 25.012& 25.279& 25.295&  25.320\\[2mm]
 \multirow{2}{*}{11}  & 25.148& 25.250& 25.169& 25.007& 24.982& 25.111& \multirow{2}{*}{1.011}\\
&24.888& 24.916& 24.898& 24.891& 24.581& 24.83 \\[2mm]
    \multirow{2}{*}{12}  & 25.312& 25.089& 25.322& 25.140& 24.813& 25.1& \multirow{2}{*}{1.019}\\
&24.753& 24.304& 24.502& 24.966& 24.751& 24.655 \\[2mm]   
    \multirow{2}{*}{13}  & 25.165& 25.203& 25.165& 25.536& 25.288& 25.271& \multirow{2}{*}{1.020}\\
&24.819& 24.874& 24.860& 24.736& 24.526& 24.763 \\[2mm]
    \multirow{2}{*}{14}  & 24.955& 24.914& 24.971& 24.739& 24.763& 24.868& \multirow{2}{*}{\textbf{1.031}}\\
&24.447& 24.038& 24.076& 23.917& 24.085& 24.113 \\[2mm]
    \multirow{2}{*}{15}  &  25.265& 25.660& 25.462& 25.343& 25.108& 25.368& \multirow{2}{*}{\textbf{1.035}}\\
&24.532& 24.716& 24.745& 24.113& 24.436& 24.509\\
    \hline
  \end{tabular}}    
  \end{center}
  \caption{The food industry example data}
  \label{tab:mueslidata}
\end{table}

\end{document}